\documentclass{aastex}

\def\ltsima{$\;\buildrel < \over \sim \;$}
\def\simlt{\lower.5ex \hbox{\ltsima}}

\begin{document}

\title{Water Vapor in Carbon-rich AGB Stars from the Vaporization of Icy
Orbiting Bodies}
 
\author{K. E. Saavik Ford \& David A. Neufeld}
\affil{Dept. of Physics \& Astronomy, The Johns Hopkins University, 3400 N. Charles St., Baltimore, MD 21218-2686}
\email{saavik@pha.jhu.edu, neufeld@pha.jhu.edu}

\begin{abstract}

We argue that the presence of water vapor in the circumstellar outflow of a
carbon-rich AGB star is potentially a distinctive signature of extra-solar
cometary systems.  Detailed models show that at suitable distances
from the star, water ice can
survive well into the carbon-rich AGB phase; water vapor
abundances as large as
 $10^{-6}$ could result from the vaporization of a
collection of orbiting icy
bodies with a total mass comparable to what might
have been originally present
in the solar system's Kuiper Belt.  In
particular, the recently-reported
 detection by the Submillimeter Wave
Astronomy Satellite of water vapor in the
 circumstellar outflow of IRC+10216
can be explained if $\sim 10$ Earth masses of ice is present at a distance
$\sim 300$~AU from that
 carbon-rich  star. Future observations with the
Herschel Space Observatory
 (HSO, formerly  known as FIRST) will facilitate 
sensitive multi-transition observations of water, yielding line ratios that can
establish the radial distribution of
water vapor in
IRC+10216.  The greater sensitivity of HSO will also allow
searches for water
 vapor to be carried out in a much larger sample of
carbon-rich AGB stars.
 
\end{abstract}


\section{Introduction}

One of the most exciting developments in astronomy during the past
decade has been the unequivocal detection of extra-solar system planets of
size as small as a few Jupiter masses (e.g.\ Mayor \& Queloz 1995,
Marcy \& Butler 1996).  This development suggests
the possibility that planetary systems around other stars harbor yet
smaller bodies, including Earth-sized planets, asteroids and comets.
The presence of comets, in particular, has been suggested by the 
$\beta$ Pictoris phenomenon, in which time-variable line absorption is
observed in a stellar spectrum at substantial Doppler shifts relative to the
central star (Ferlet, Vidal-Madjar, \& Hobbs 1987),
suggesting comets evaporating as they speed near the central star.
The discovery of extra-solar planetary systems also raises the question
of how such systems  change as the stars that they surround evolve.  

Stern, Shull \& Brandt (1990; hereafter SSB90) have pointed out 
when a star evolves off the main sequence, the resultant increase
in its luminosity must lead to the vaporization of any 
icy bodies orbiting within several hundred AU, an effect that could result
in the release of significant amounts of water vapor
if the star were surrounded by a Kuiper Belt or Oort cloud.
SSB90 considered the vaporization of icy bodies of various radii at
various distances from a post-main-sequence star of constant assumed
luminosity $6000\,\rm L_\odot$,
and computed the resultant rate at which water vapor is deposited in
the circumstellar environment.  They suggested that this process 
could be responsible for the large water vapor and OH abundances
typically observed in the circumstellar envelopes of oxygen-rich stars.

Although large water vapor abundances are in any case expected
around oxygen-rich stars, and although the water outflow rates
around many such sources greatly exceed what can plausibly be 
explained by the SSB90 model, recent measurements of modest water vapor
abundances in two {\it carbon-rich} circumstellar outflows (Herpin \&
Cernicharo 2000, Melnick et al. 2001) motivate
renewed attention to the mechanism proposed by SSB90.
In particular, Melnick et al.\ have used the
{\it Submillimeter Wave Astronomy Satellite} (SWAS) to detect 
water vapor emission from the classic carbon star IRC+10216.
Carbon stars, which are late-type AGB stars where the
carbon-to-oxygen ratio is greater than $1$, are expected
to harbor almost no water vapor in their circumstellar environments;
the equilibrium chemistry of
oxygen is dominated by CO and there is ordinarily very little oxygen left to
form any other molecules.   The average water abundance predicted by standard
chemical models for IRC+10216 (Millar et al.\ 2000) is less than $10^{-12}$
(Millar \& Herbst 2001) 
and thus the presence of detectable water
vapor in IRC+10216
 may be a distinctive signature of the vaporization of
orbiting icy bodies.

The existence of this signature rests critically
upon
the survival of such bodies.  In particular, it is not
possible to determine from the study of SSB90 whether {\it any}
icy bodies  will survive the $\sim 1$~Gyr 
of post-main-sequence evolution that precedes the carbon-rich
phase (at least at radii where such bodies are plausibly
present in the first place and where they 
would be subject to vaporization by
a carbon-rich AGB star). To address this question
we have extended the work of SSB90 by considering (1)
the exact
 luminosity variations expected in a post-main-sequence star; 
(2) the size distribution and total mass of icy bodies expected 
if extra-solar
cometary systems are similar to our own Kuiper
Belt (the properties of which have been
greatly elucidated [e.g.\ Jewitt \& Luu 2000] since
the work of SSB90); and (3) the
specific application to carbon-rich stars, where the vaporization of icy
bodies has the most distinctive observational signature.
Our paper is organized as follows: in \S 2, we detail our theoretical
model of the vaporization of a Kuiper Belt analog  around a post-main-sequence
star; in \S 3 we present the results of our calculations; the results are
discussed in \S 4.

\section{Calculations}

For our standard model, we considered a star of mass $M_*=
1.5 M_{\odot}$ and followed its luminosity evolution from
the main sequence to the late AGB stage using evolutionary tracks kindly
provided by Allen Sweigart.  The luminosity, $L$, during the last $\sim
2\,$Myr of the AGB track (the thermal pulsation or ``TP-AGB" phase) is shown
in
Figure 1; the zero of time corresponds to the first thermal pulse on
the AGB. The stellar evolution model assumes a 
mass-loss rate, ${\dot M}_{*}$, given by the
expression of Reimers (1975), $({\dot
M}_{*}/M_{\odot} \rm yr^{-1})
  = 4 \times 10^{-13}\it \eta_{R}
(L/L_{\odot})(R_*/R_{\odot})(M_*/M_{\odot})^{-1}$, where $R_*$ is the
stellar radius, and with
the ``Reimers parameter''$\eta_R$ equal to 0.4.  With this prescription
for the mass-loss rate, the stellar mass has decreased to $1.4\,M_{\odot}$
by the start of the AGB phase and to $0.92\,M_{\odot}$ by the 
end of the AGB phase.

Using this model for the luminosity evolution of a $1.5M_{\odot}$ star, 
we have calculated how a population of orbiting icy bodies would evolve. 
The mass loss rate per unit surface area, $\dot m(T)$, is described
by the equation
$\dot m(T) = P_{s}(T) (\mu/ 2 \pi kT)^{1/2}$, 
where $\mu$ is the molecular mass of water, $k$ is the Boltzmann 
constant and $P_s(T)$ is the vapor pressure of water ice at temperature $T$. 
This equation, which neglects gravitational effects,
applies to bodies of radius $\simlt 1000$~km for which the escape
velocity is smaller than the typical outflow velocity  of the vaporizing  
molecules ($\sim 1 \, \rm km \,s^{-1}$).  Furthermore, it applies
strictly to bodies composed of pure water ice rather than to bodies 
composed of a mixture of ice and refractory material.  In large
icy bodies (the size of Pluto and Charon), it is very plausible that
differentiation will lead to a structure in which relatively pure
ice surrounds a rocky core (McKinnon, Simonelli \& Schubert 1997).  In smaller
bodies, however, the dust and ice will be well mixed and the effects of
the refractory material on the vaporization rate is 
less clear -- and may depend critically upon the size distribution
of the dust (Prialnik 1992; Orosei et al.\ 1995).
The average surface
temperature in equilibrium, $T$,  reached by an icy body at distance
$R$ from a star of luminosity $L$ is given by
$\epsilon \sigma T^4 = [(1-A) L/(16\pi R^2) - H {\dot m}(T)]$, 
where $A$ is the albedo, $H= 2.45\times 10^{10} \rm
\,erg\,g^{-1}$ is the specific heat of vaporization (Prialnik 1992),
$\epsilon$ is the
 emissivity and $\sigma$ is the Stefan-Boltzmann constant.  
Following SSB90, we adopt an albedo $A=0.04$, an emissivity $\epsilon = 1$,
and a vapor pressure $P_{s}(T) = 10^{9.183837 - 2403.4/T} \,\, {\rm
torr}$ (Lebofsky 1975).

The mass-loss caused by vaporization reduces the radius of
each orbiting icy body.
 For a spherical body of uniform 
density $\rho$, the
change
 in radius after time $t$ is given by 
$\Delta r(t) = \int_{0}^{t} ({\dot m(t') /\rho})\, dt'$.
After time $t$, all objects of initial radius smaller
than $\Delta r$ have been completely vaporized.
In computing the total mass loss rate from a population of
orbiting icy bodies, we assume an initial size distribution of the
form  $dn/dr \propto r^{-q}$, where $dn$ is the number of objects of
radius $r$ to $r+dr$.
Recent observations (Jewitt \& Luu 2000) of our own Kuiper Belt suggest
$q \sim 4$ for radii $r$ between
$r_{\rm min} \sim 1\,$km and $r_{\rm 
max} \sim 1200\,$km  (the size of Pluto);
in other words, the size distribution has equal amounts
of ice mass per logarithmic radius interval and the 
available reservoir of ice is dominated neither 
by small nor by large bodies.  Although well-motivated
by observations of the past decade, our assumption that
the Kuiper belt contains objects with a range of sizes
is not a critical one, and calculations with a single
assumed object size yield qualitatively similar results. 

\section{Results}

Our calculation shows that at the astrocentric radii
of the classical Kuiper Belt, $R = 30 - 50 \rm \,AU$ (Jewitt \& Luu
2000), even large icy bodies are entirely vaporized prior 
to the TP-AGB phase\footnote{Note, however, that any icy bodies
{\it originally} at the radius of the classical Kuiper Belt may
have experienced significant outwards migration during the
post-main-sequence evolution as a result of stellar mass-loss.}.  
Thus by the time the
dredge-up of carbon
 has made the photospheric C/O ratio greater 
than 1, the release
 of water vapor can only take place if icy bodies are
present
 at larger radii.  Figure 2 shows the ``complete
vaporization radius'', 
 $R_{\rm vap}$, 
within which every icy body has been destroyed, 
as a function of time after the start of the TP-AGB phase.
The solid curve shows the results for a size distribution
extending to $r_{\rm max} = 1200\,$km,
the dotted curve  
for the case $r_{\rm max} = 120\,$km, and the dashed curve 
for the case $r_{\rm max} = 12\,$km.
Clearly, the complete vaporization radius is a slowly increasing
function of $r_{\rm max}$, because larger bodies have a
smaller surface-area-to-mass ratio and thus a longer lifetime.
 
Figure 3 shows the evolution of the water outflow rate and abundance
for a collection of bodies located in circular
orbits at fixed distance $200\,$AU from the central star.  The mass-loss
rate per unit initial ice mass, 
$\dot M({\rm H_2O})/ M_0 ({\rm ice})$,
is shown in Figure 3a for the case 
$r_{\rm max} = 1200\,$km.  The quantity $M_0 ({\rm ice})$
refers to the total mass of water ice present in
the collection of orbiting bodies prior to the onset
of evaporation.  The same results
appear in Figure 3b, but are now expressed as a water abundance,
$x({\rm H_2O}) \equiv n({\rm H_2O})/n({\rm H_2})$.  
Given a water mass outflow rate of $\dot M({\rm H_2O})$, the water 
abundance in the outflowing gas is 
${1 \over 9} \times {\dot M({\rm H_2O})}/{3 \over 4} \times 
{\dot M_{*}}$ relative to H$_2$, the factor of  
${3 \over 4}$ being the mass fraction 
of hydrogen and the factor of ${1 \over 9}$ being the ratio of 
the H$_2$ molecular mass to that of H$_2$O. 
Because the
H$_2$ mass-loss rate is proportional to the Reimers parameter,
$\eta_R$, the quantity plotted in Figure 3b is
$x({\rm H_2O}) \eta_R / M_0 ({\rm ice})$.
We note that our use of the
Reimers parameter $\eta_R$ in Figure 3b is
merely a useful way of parameterizing the mass-loss rate 
and that the results presented here apply
{\it regardless} of whether ``Reimers Law'' is correct -- i.e.\ regardless of
whether  $\eta_R$ is a constant.  Indeed, there is a developing
consensus (e.g. Bl\"ocker 1995, Willson 2000) that $\eta_R$
is an increasing function of luminosity and that mass-loss 
occurs preferentially near the tip of the AGB. The current value of
$\eta_R$ for the high mass-loss star IRC+10216 is $\sim 10$.

In Figures 4a and 4b, the water outflow rate and abundance
are now shown {\it as a function of astrocentric distance}, $R$, 
at the three example evolutionary stages marked by open symbols on Figures 1 
and 3.  
As in Figure 3, these results apply to
a collection of icy bodies all located at a single distance, $R$,
from the star.  At any given time,
the water outflow rate per unit initial ice mass
shows a peak at some particular distance.
At small distances from the star, the outflow rate and abundance 
increase as the assumed distance gets larger because the amount
of icy material remaining is an increasing function of $R$.  (This
increase is a gradual one because
we assume the presence of a range of sizes for the orbiting
icy bodies.)  At large astrocentric distances, the outflow rate
and abundances drop with increasing $R$ because of the diminishing
stellar flux.
The radius at which  
$\dot M({\rm H_2O})/ M_0 ({\rm ice})$ peaks moves
outwards as the star evolves because of the steady 
increase in stellar luminosity on the AGB.
Figure 4a implies that for icy bodies at the
right distance from the star, 
water outflow rates as large as $\sim 10^{-5} M_0 ({\rm ice})$~yr$^{-1}$ 
can plausibly be attained during the TP-AGB phase.

\section{Discussion}

The results presented above can now be discussed in the context
of recent SWAS observations of the carbon-rich AGB star IRC+10216.
The extraordinary stability of the SWAS receivers has allowed
radiometric performance to be achieved in observations of duration
as long as 200 hours (Melnick et al.\ 2000).  Long duration observations have
thereby  achieved sensitivities considerably better than had been envisaged
prior to the launch of SWAS.  Such observations have led to the
unexpected detection of the $1_{10}-1_{01}$ 557 GHz transition 
of water vapor in emission toward IRC+10216 (Melnick et al.\ 2001;
hereafter M01).  The measured line strength implies a water abundance
in the outflowing circumstellar envelope in the range 4 -- 24 
$\times 10^{-7}$ (M01) 
corresponding to a water mass-loss rate of $2 - 4\times 10^{-10}
M_\odot$~yr$^{-1}$ $= 0.6 - 1.4 \times 10^{-4} M_\oplus$~yr$^{-1}$.  
These exact
values depend upon the H$_2$ mass-loss rate and the distance to IRC+10216,
both of
which are somewhat uncertain.

The exact initial mass of IRC+10216 is poorly-known
but is believed to be between $1.5$ and $4\, M_{\odot}$
(Forestini, Guelin \& Cernicharo 1997, Kahane et al. 2000).
We have carried out calculations analogous to those
presented in \S 3 with a stellar mass of $4\, M_{\odot}$  
in place of the standard case $M_* = 1.5 \, M_{\odot}$;
the results are qualitatively similar to those for the
standard case, implying that our results are neither
critically dependent upon the mass of the star nor upon
the exact (and somewhat uncertain) details of the luminosity
evolution model.

The results shown in Figure 4a and discussed in \S 3 above 
imply that the vaporization of icy bodies can explain
the observed abundances of water vapor in the IRC+10216
outflow, given a total initial ice mass, $M_0({\rm ice})$,
that can be as little as $10\,M_\oplus$ if the
icy bodies are located at a suitable radius.  This value
increases to $\sim 50\,M_\oplus$ in a distributed
model in which the icy bodies are assumed to be
spread over a broad range of
radii (40 -- 400 AU) with the surface density of 
icy bodies proportional to $R^{-2}$. 
The exact value
of $M_0({\rm ice})$ required depends upon several uncertain
parameters, including the actual water abundance, 
the range of astrocentric radii at which 
icy bodies are present, the maximum object radius, 
and the evolutionary stage\footnote{From 
the fact that IRC+10216 possesses one of the highest
mass-loss rates of any known carbon star -- a mass-loss rate
some two orders of magnitude above the median value for
carbon AGB stars -- it seems likely that this source
is in a stage of enhanced luminosity close to the tip of
the AGB.} of the star.

Although the properties of any Kuiper Belt analog
around IRC+10216 might be quite different from those
of our own Kuiper Belt, it is of interest to compare
the required value of $M_0({\rm ice})$ with the
mass inferred for the one system of which we have 
direct knowledge.   Dynamical considerations place a limit of $\sim
1\,M_\oplus$ upon the mass located at heliocentric radii in the range
30 -- 50~AU (i.e.\ in the "classical Kuiper belt"), 
while recent detections of large Kuiper Belt
Objects (KBOs) suggest a total
mass $\sim 0.1\, M_\oplus$
 (Jewitt 1999), of which $\sim 50\%$ is likely
water ice. 
Thus the observed amount of water ice currently present
in the nearby classical Kuiper belt is far below what would be needed to
account for the water abundance observed in IRC+10216.
However, models for the formation of large KBOs such as
Pluto (Kenyon \& Luu 1998), suggest that the Kuiper Belt mass must
$\it originally$ have been 
at least $10 \, M_\oplus$ during the
early history of
 the solar system; thus the vast majority of the material
has likely been ejected to larger heliocentric distances in the Kuiper
belt or Oort cloud where it cannot be detected in current surveys.

In addition to water vapor, the vaporization of comet-like icy bodies
could release a distinctive mixture of other oxygen-bearing molecules 
that are not expected in carbon-rich environments but which are
typically observed in comets.  These include carbon dioxide and
methanol. 
We are not aware of any detections
of these molecules\footnote{Latter \& Charnley (1996) reported the detection
of millimeter line emission at the frequencies of several methanol
transitions, but subsequently favored C$_4$H and C$_4$H$_2$ as the
species responsible for the emission originally attributed to 
methanol.} toward IRC+10216 nor of any upper limits that 
provide useful constraints upon their abundances.

The discussion presented above suggests that the
vaporization of orbiting icy bodies is a plausible explanation
for the water vapor detected in the circumstellar environment
of IRC+10216.  
In addition, the vaporization scenario makes a specific 
prediction for the spatial distribution of the circumstellar
water vapor.  As the results
in Figure 2 show, the vaporization of icy bodies injects water
vapor into the outflow only at radii greater than
$\sim 75$~AU.
  As noted by M01, the strength of the 
low-lying submillimeter transition
observed by SWAS relative to higher-excitation
far-infrared transitions will be higher in the
vaporization scenario than would be the case were the water vapor abundance
uniform throughout the envelope.  While upper
limits on 
far-infrared water line luminosities obtained by the
{\it Infrared Space Observatory} (Cernicharo et al.\ 1996) provide
marginal evidence against a uniform water abundance 
(M01), 
definitive observations of the water
line ratios will have to await the
greater sensitivity of the 
{\it Herschel Space Observatory} (HSO, formerly known as FIRST).
The expected sensitivity of HSO -- probably three orders of magnitude
better than SWAS for observations at 557 GHz -- will also allow
searches for water vapor to be carried out in a much larger sample of
carbon-rich AGB  stars.

\acknowledgments
We thank Allen Sweigart for providing us with evolutionary
tracks of 1.5 and 4~$M_\odot$ stars.  We gratefully acknowledge helpful
discussions with Gary Melnick, Eric Herbst, Al Glassgold
and David Hollenbach.
This work was
supported by subcontract SV252005 from the
Smithsonian Institution.

\clearpage
\centerline{\bf Figure Captions}
\parindent 0pt

Fig. 1 -- Evolutionary tracks for a 1.5 $M_\odot$ star (kindly provided
by Allen Sweigart).  The luminosity, $L$, during the last $\sim 
2\,$Myr of the AGB track (the thermal pulsation or ``TP-AGB" phase) is shown
in Figure 1; the zero of time corresponds to the first thermal pulse on 
the AGB.  Open symbols mark
the three example evolutionary stages plotted in Figure 4.

Fig. 2 -- Vaporization radius, 
$R_{\rm vap}$,  
within which every icy body has been destroyed, 
as a function of time after the start of the TP-AGB phase. 
The solid curve shows the results for a size distribution 
extending to $r_{\rm max} = 1200\,$km, the dotted curve 
for the case $r_{\rm max} = 120\,$km, and the dashed curve 
for the case $r_{\rm max} = 12\,$km

Fig 3. -- Evolution of the water outflow rate and abundance 
for a collection of bodies located in circular orbits at  
fixed distance $200\,$AU from the central star.  The mass-loss 
rate per unit initial ice mass,  
$\dot M({\rm H_2O})/ M_0 ({\rm ice})$, 
is shown in Figure 3a for the case  
$r_{\rm max} = 1200\,$km.  The quantity $M_0 ({\rm ice})$ 
refers to the total mass of water ice initially present in 
the collection of orbiting bodies.  The same results 
appear in Figure 3b, but are now expressed as a water abundance, 
$x({\rm H_2O}) \equiv n({\rm H_2O})/n({\rm H_2})$. Open symbols mark the
three example evolutionary stages plotted in Figure 4.

Fig 4. -- Water outflow rate and abundance 
{\it as a function of astrocentric distance}, $R$, 
at the three example evolutionary stages marked by open 
symbols on Figures 1 and 3.  The short-dashed line refers to 
the evolutionary stage marked by a triangle in Figures 1 and 3,
the long-dashed line to that marked by a diamond, and the solid line
to that marked by a star.  
As in Figure 3, the results in Figures
4a and 4b apply to
a collection of icy bodies all located at a particular distance, $R$,
from the star.

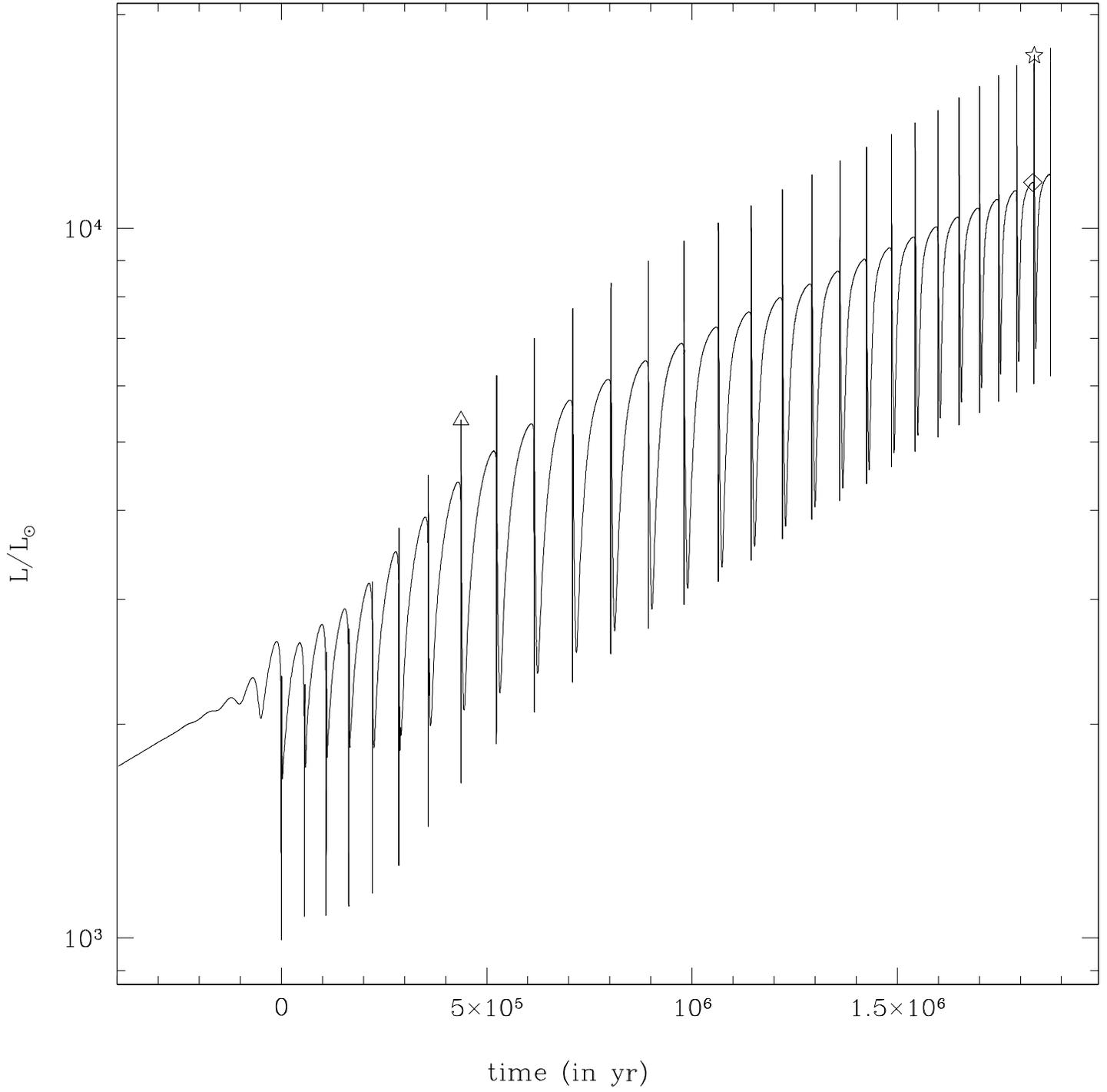
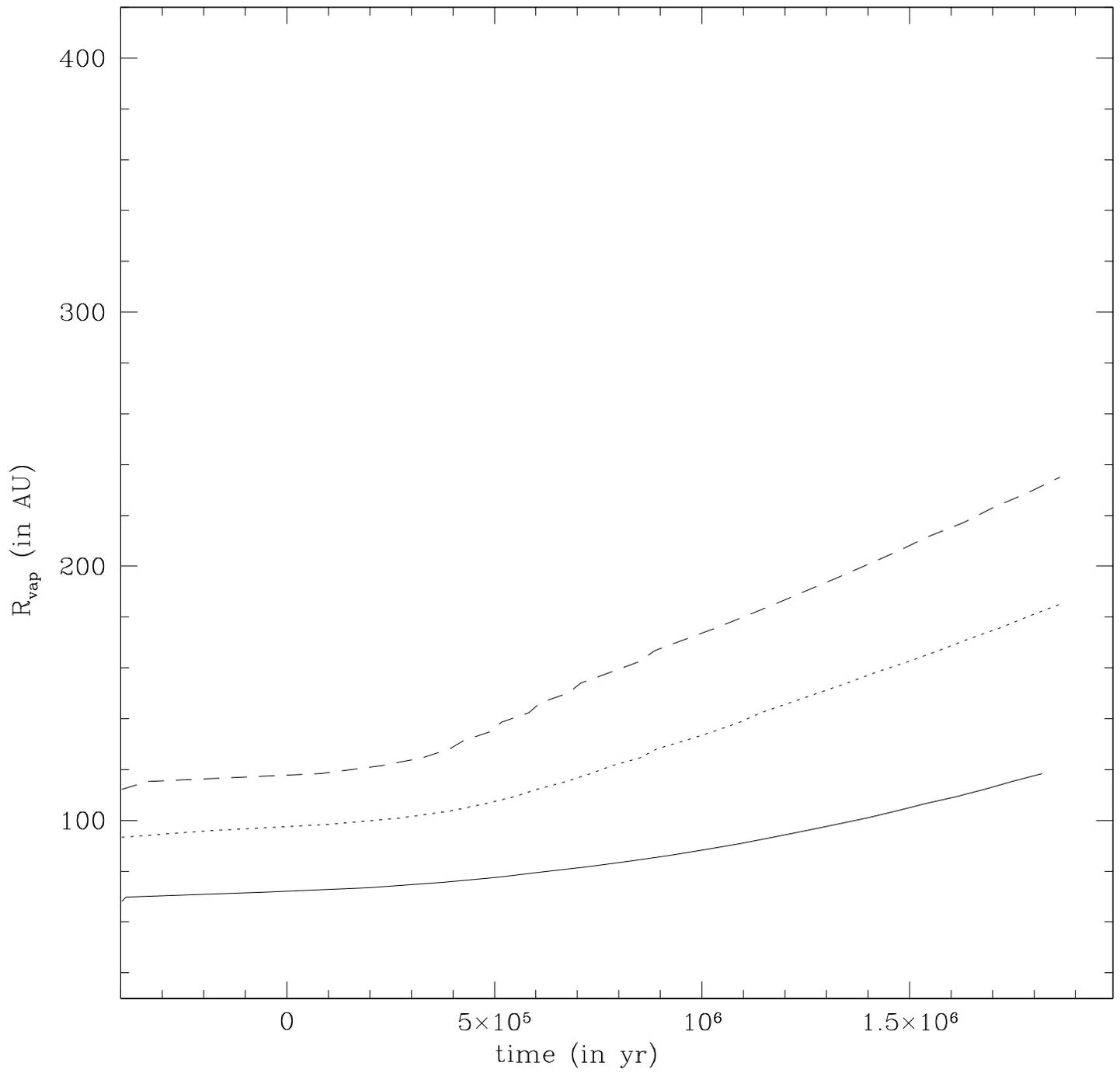
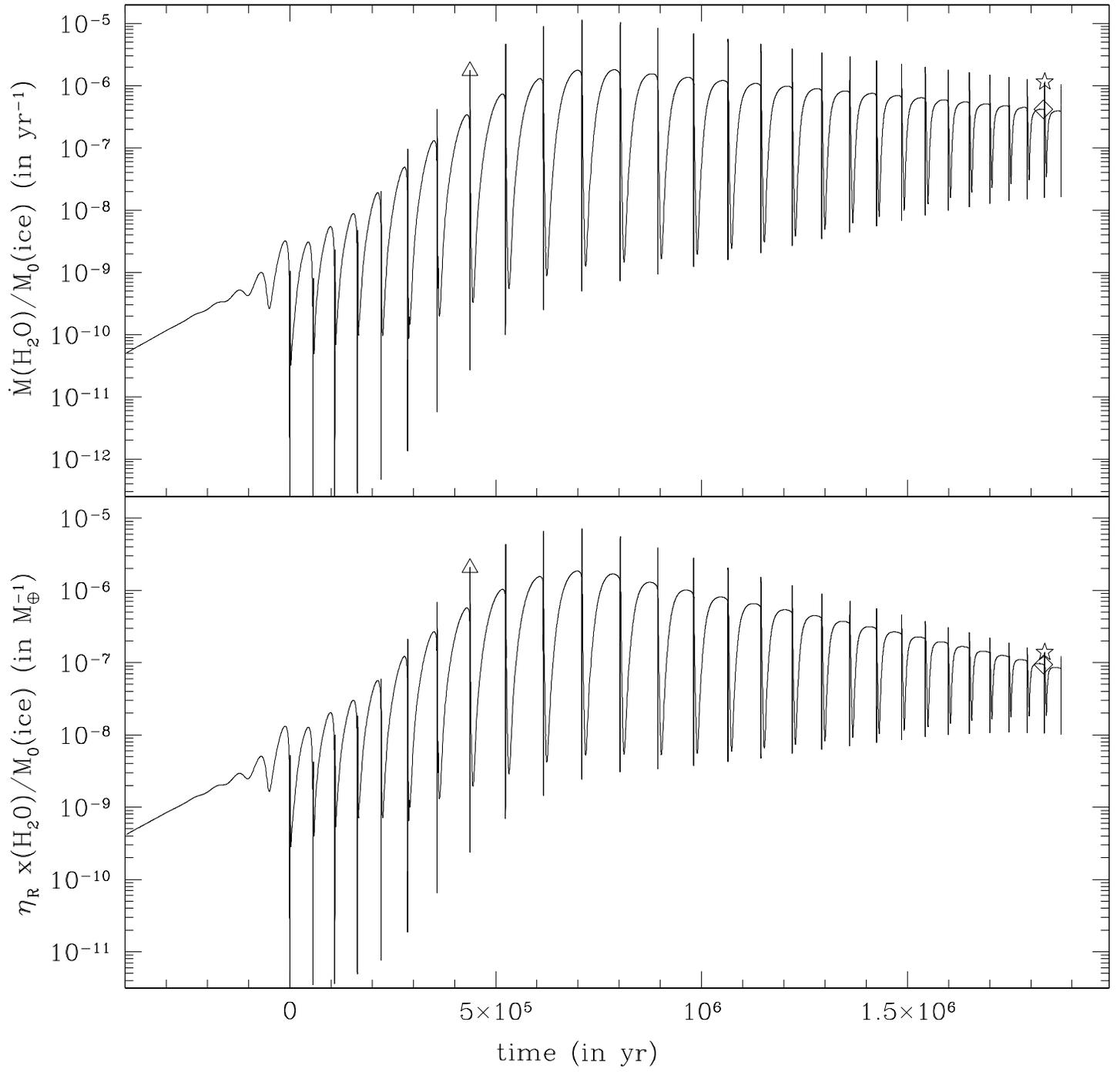
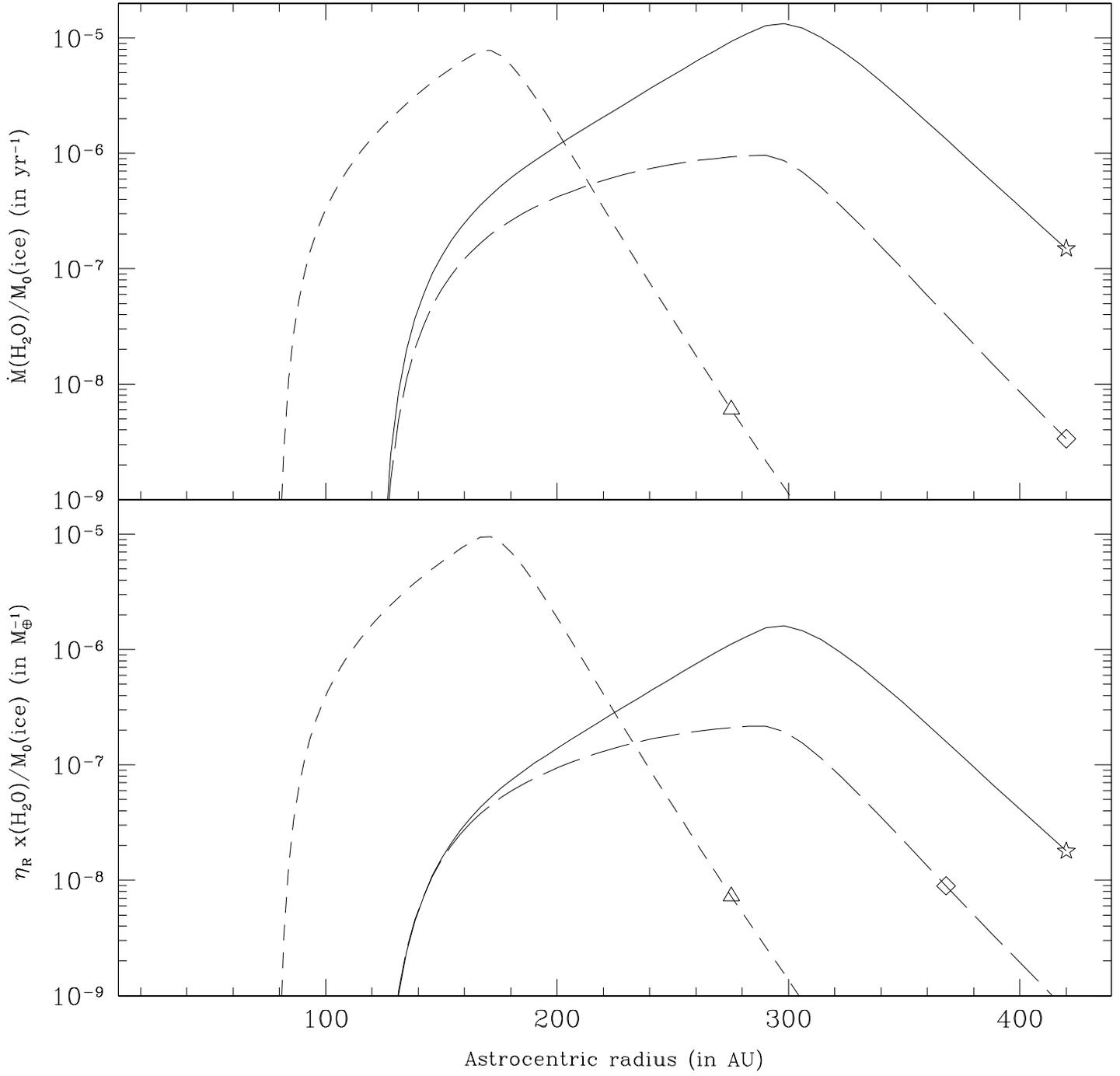
\end{document}